\documentclass[11pt]{article}

\usepackage[utf8]{inputenc}
\usepackage[T1]{fontenc}
\usepackage{helvet}
\usepackage{setspace}
\usepackage{verbatim, authblk}
\usepackage{ulem}
\usepackage{graphicx, subcaption} % For graphics, we recommend your use graphicx.sty.
\usepackage{textcomp}
\usepackage{array}
\usepackage{makecell}
\usepackage[a4paper, top=2cm,bottom=2cm,left=0.7in,right=0.7in,marginparwidth=1.2cm]{geometry}
\usepackage{xcolor}
\usepackage{amsmath,bm}
\usepackage{float}
\usepackage{stmaryrd}
\usepackage{comment}

\usepackage{xr}
\makeatletter
\newcommand*{\addFileDependency}[1]{% argument=file name and extension
\typeout{(#1)}% latexmk will find this if $recorder=0
\@addtofilelist{#1}
\IfFileExists{#1}{}{\typeout{No file #1.}}
}\makeatother

\newcommand*{\myexternaldocument}[1]{%
\externaldocument{#1}%
\addFileDependency{#1.tex}%
\addFileDependency{#1.aux}%
}
\myexternaldocument{Supplementary_information}
\usepackage[square,numbers,sort]{natbib}

    \newcommand{\Int}[1]{\begin{aligned} \int_{#1} \;\end{aligned}}
    
    \newcommand{\OInt}[1]{\begin{aligned} \oint_{#1} \;\end{aligned}}
    
    \newcommand{\Id}{\mbox{$\underline{\underline{I}}$}} % identité

\newcommand{\stress}{\mbox{$\underline{\underline{\sigma}}$}}

\newcommand{\D}{\mbox{$\underline{\underline{D}}$}} % Taux de déformation
 \title{Using lateral dispersion to optimise microfluidic trap array efficiency} 

\author[,1]{Nicolas Ruyssen\thanks{Corresponding author: \texttt{nicolas.ruyssen@sorbonne-universite.fr}}}
\author[2, 3]{Gabriel Fina}
\author[4, 5]{Rachele Allena}
\author[6]{Marie-Caroline Jullien}
\author[2,3]{Jacques Fattaccioli}

\affil[1]{Sorbonne Université, Physico-Chimie des Electrolytes et Nanosystèmes Interfaciaux, PHENIX, CNRS UMR 8234, Paris F-75005, France}
\affil[2]{PASTEUR, Département de Chimie, École Normale Supérieure, PSL Université, Sorbonne Université, CNRS, 75005 Paris, France}
\affil[3]{Institut Pierre-Gilles de Gennes pour la Microfluidique, 75005 Paris, France}
\affil[4]{Université Côte d’Azur, Laboratoire Jean Alexandre Dieudonné UMR CNRS 7351, Nice, France}
\affil[5]{Institut Universitaire de France}
\affil[6]{Univ. Rennes, CNRS, IPR (Institut de Physique de Rennes) UMR $6251$, F-$35000$ Rennes}

\date{}
\begin{document}
\raggedbottom
\maketitle


\smallskip

\begin{abstract}
 Microfluidic trapping arrays have proven to be efficient tools for various applications that require working at the single-cell level, such as cell-cell communication or fusion. Although several hydrodynamic trapping devices have already been optimised, two-dimensional (2D) single-layer trapping arrays with high trap densities remain partially inefficient. Specifically, many traps remain empty, even after prolonged injection, which drastically reduces the number of samples available for post-treatment. These unfilled traps result from the symmetrical nature of the flow around the traps, and breaking this symmetry enhances capture efficiency. In this study, we use a numerical approach to show that optimal geometries can significantly increase filling efficiency and a preliminary  experimental test confirming our approach is provided. We show that these improvements are achieved by promoting lateral dispersion of particles, facilitated either through an optimised oblique flow or by introducing disorder into the spatial arrangement of traps without specific inlet/outlet adjustment.

\end{abstract}

\newpage
\section{Introduction}
Microfluidic trapping arrays are of significant interest due to their capacity to precisely capture and manipulate individual cells or particles within a highly controlled environment. These arrays enable high-throughput analysis and long-term observation, both essential for advancing research in cell biology, drug screening, and diagnostics. Over the years, a broad range of technical solutions has been developed: trapping can be achieved passively through gravity \cite{Charnley2009, Fradet2011, Figueroa2010, Rousset2017}, inertial lift \cite{Rousset2017}, or permanent magnets \cite{Winkleman2004, Smistrup2006}, as well as through active methods such as valves \cite{Zhou2016, Au2011}, optical tweezers \cite{Grigorenko2008}, dielectrophoresis \cite{Rosenthal2005, Challier2021}, electromagnets \cite{Lee2004}, or acoustics \cite{Evander2007}. Among these methods, hydrodynamic trapping arrays are the simplest, operating by suspending objects in a fluid that flows through a microfluidic chamber containing an array of traps. Since the objects targeted for immobilization follow the fluid flow, they are captured in these traps, either individually or in quantities that depend on the size ratio between the target objects and the traps. Thanks to their apparent simplicity in implementation and relative ease in terms of microfabrication, the literature attached to hydrodynamic arrays is vast.

The optimisation of trapping efficiency relies on tailoring the flow inside the chamber which goes through the trapping array. The most common hydrodynamic trapping arrays so far consist of two-dimensional (2D) systems with traps arranged in a regular staggered pattern within a Hele-Shaw cell \cite{Skelley2009, Pinon2022a, Pineau2022a, Wlodkowic2009, DiCarlo2006}. Although several hydrodynamic trapping devices have been optimised, trapping arrays made by single-layer lithography remain only partially efficient. Indeed, many traps frequently remain empty even after prolonged injection, which significantly reduces the number of samples available for further analysis \cite{Wlodkowic2009}. Trapping efficiency improves with double-layer lithography techniques \cite{DiCarlo2006, Skelley2009, Pinon2022a}, though this method complicates fabrication.

Recently, we have demonstrated that both the efficiency and trapping homogeneity of microfluidic arrays, fabricated via single-layer lithography, can be significantly improved by tailoring the flow within the chamber through precise inlet and outlet positioning \cite{Mesdjian2021}. This configuration directs the carrier fluid to flow diagonally relative to the chamber’s main axis, inducing flow asymmetry in front of each trap, which enhances particle immobilization probability. Additionally, most numerical studies rely on 3D \cite{Rousset2017,Kobel2010,SohrabiKashani2019,Wlodkowic2009,Wlodkowic2010} or 2D with explicit solid particles \cite{Xu2013_mef,Xu2020,SohrabiKashani2019} models which both are too computationally intensive for an optimisation study of dense trapping arrays (of approximately $100$ traps). we have achieved strong agreement with experimental results using a simplified 2D Computational Fluid Dynamics (CFD) approach, enabling an increase in the number of geometries studied with minimal computational cost \cite{Mesdjian2021}.

In this study, we investigate how trapping efficiency and homogeneity can be optimised. Firstly, we optimise trapping efficiency independently across five dimensionless geometric parameters. Secondly, we demonstrate by simulation that comparable trapping efficiencies can be obtained with disordered trapping arrays and experimentally verify the potential of such geometries for one example. Finally, we conduct a quantitative analysis of the lateral dispersion of Lagrangian particles within the optimised geometries. This analysis confirms the critical role of lateral dispersion in facilitating large-scale particle trapping. These comprehensive results enable us to identify the most important parameters to consider for achieving efficient trapping arrays and to gain a deeper understanding of the underlying mechanisms that enhance trapping efficiency.

\section{Materials and methods}
\label{section_MM}

\subsection{Parameterised geometry}
\begin{figure}[H]
    \centering
    \includegraphics[]{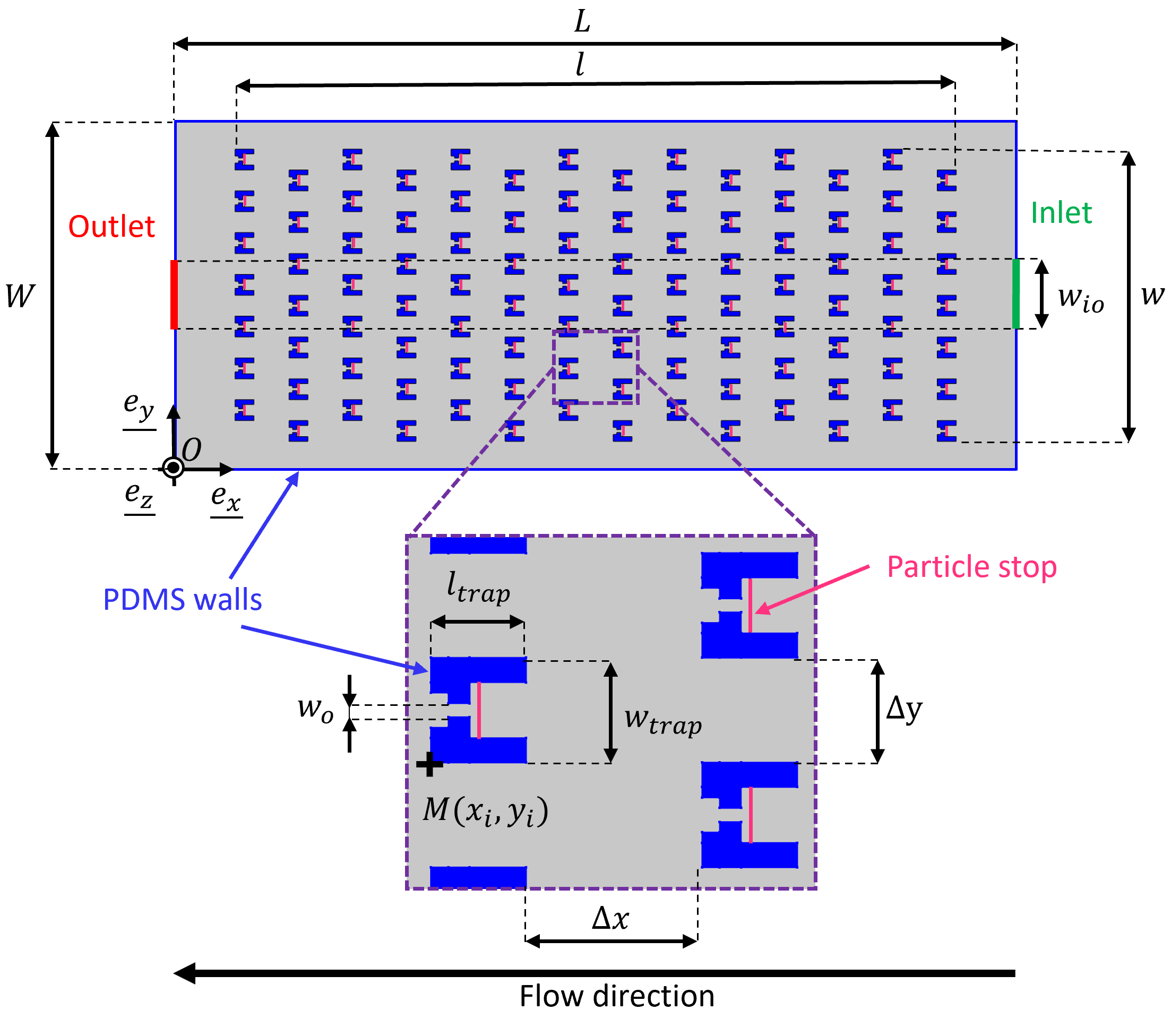}
    \caption{The model geometry. The fluid domain is represented in grey, the traps and the PDMS walls in blue, the inlet resp. outlet channels in green resp. in red. A particle stop frontier is added in each trap (in pink) to virtually stop particles without modifying the fluid flow.}
    \label{fig_geom}
\end{figure}

The whole numerical model is developed using the commercial Finite-Element software COMSOL Multiphysics \textregistered \ while post-processing is achieved with Python. We study a single layer trapping array of $98$ traps as previously \cite{Mesdjian2021}. Because of symmetry, a $2$D model built from parameterised geometric primitives is suggested (Figure \ref{fig_geom}). The trapping array is initially organised in a rectangular staggered pattern of axis $(Ox)$ in the streamwise direction and of axis $(Oy)$ in the lateral direction, (see Figure \ref{fig_geom} where all the geometrical parameters were sketched). The dimensions of the trapping array are $l$ and $w$ and the one of the cavity are $L$ and $W$ in the streamwise resp. lateral directions; and the distances between the traps are respectively $\Delta x$ and $\Delta y$. We call "chamber" the whole system including the trapping array and the cavity. For more information, all the fixed geometric parameters of the model are reported in supplementary Table \ref{tableau_all_param} section \ref{sec_SI_param_values}. 

\subsection{Physics of the problem}

We consider the flow as incompressible, stationary and the fluid as Newtonian. Furthermore, the maximal Reynolds number $Re=\frac{Uh}{\nu}$ (where $U \sim 1$ mm s$^{-1}$ is the maximal mean velocity, $h = 14$ µm the height of the cavity and $\nu = 10^{-6}$ m$^2$ s$^{-1}$ the kinematic viscosity of the solution) is about $Re = 0.01 $, which allows us to consider the Stokes equations. In order to simplify the study, the fluid flow is reduced in 2D in the in-plan of the cavity (\textit{i.e.} $(O,\underline{e_x}, \underline{e_y})$, Figure \ref{fig_geom}) and verifies the Brinkman equations that account for both viscous dissipation in the in-plane and in the out-of-plane direction of the cavity (\textit{ie} $(O,\underline{e_z})$, Figure \ref{fig_geom}) through a body force $\underline{f}$. As such, the latter equations read: 
\begin{equation}
\begin{cases}
    - \boldsymbol{\nabla}p + \mu \nabla^2 \underline{v} + \underline{f} = \underline{0} \\
    \boldsymbol{\nabla}  \cdot  \underline{v} = 0 
\end{cases}
    \label{eq_stokes_incomp}
\end{equation}
where $p$ is the pressure field, $\mu$ the fluid dynamic viscosity, $\underline{v}$ the velocity field. The chamber is horizontally orientated such that gravity does not contribute to the flow. The friction body force $\underline{f}$ is given in the shallow channel approximation by: 
\begin{equation}
    \underline{f}= - \frac{12 \ \mu}{h^2} \ \underline{v} 
    \label{eq_ssa}
\end{equation}
Therefore, $\underline{v}$ corresponds to the depth-averaged velocity field in-plane (in 2D). As boundary conditions, uniform pressures were applied on the inlet $p_{in} =100 $ Pa and outlet $p_{out} =0 $ Pa boundaries. On the PDMS walls, the viscous fluid-solid interaction forces a zero fluid velocity $\underline{v}  = \underline{0}$. First, equations \ref{eq_stokes_incomp}, \ref{eq_ssa} and the latter boundary conditions are numerically solved to calculate the stationary flow $\underline{v}$. To do so, these equations are translated into a mixed weak formulation and are discretised in space by applying the Finite-Element Method (FEM) with linear Lagrange interpolation functions for both pressure and velocity fields. For more information see supplementary section \ref{sec_SI_FEM}. \\
We study a dispersion of polystyrene (PS) particles of diameter $D_p = 5$ µm and of concentration $ C_p = 10^6$ ml$^{-1}$. This concentration is sufficiently small that particles do not modify the viscosity of the suspending fluid and leads to a typical distance between particles about $62$ µm which allows to neglect particle-particle interactions. Furthermore, the particles diffusion coefficient is about $8.7 \ 10$$^{-14}$ m$^2$ s$^{-1}$ which makes particle diffusion negligible with respect to particle convection. Despite a low Reynolds number, the Lifto-Diffusif number was ${\cal N} = \frac{27\pi V^2\rho R_p^4}{2k_b T h} \sim 30$, thus, lift forces are non-negligible in our flow conditions \cite{Mottin2021Nov}. However, the corresponding maximal depletion thickness at the outlet of the chamber is $\delta = \sqrt{\frac{R_p \rho L }{W h^2 \mu}} \sim 1.3 $ µm which is small compared the particle size of $5$ µm, such deviations is thus disregarded in the following. The PS particles density of $1.05$ g ml$^{-1}$ is very close to the carrier fluid one, thus particle sedimentation is also disregarded. As the Reynolds number is low and the particles are three times smaller than the channel height, we make the assumption that the particle center of mass trajectories were similar to the carrier fluid streamlines (neglected size). The particles equation of motion is obtained considering a purely advective transport:
\begin{equation}
    \frac{\partial \underline{r} }{\partial t} = \underline{v} 
    \label{PFD_point}
\end{equation}
where $\underline{r}$ is the particle position. When a particle enters into a trap, it is artificially stopped at a stop boundary in each trap (pink boundaries Figure \ref{fig_geom}). Experimentally, it is impossible to perfectly control the particle initial positions at the inlet. Therefore, we try to sweep the most possible every  positions on the inlet by regularly placing $ N_p = 200 \ (\approx 2 \times \, \rm{number \ of \ traps}$) particles on the inlet boundary. Equation \ref{PFD_point} is discretised in time and integrated using an implicit generalised-alpha numerical scheme.

\subsection{Geometry parametric analysis}
\label{sec_MM_param}
From a purely applications perspective, the reader can refer to sections \ref{sec_resultats_opti} outlining the optimum characteristics of a high-performance trapping array. This section presents the definition of the different parameters that are used for the optimisation of the trapping array. We first have to define the objective of the optimisation. As several particles can be captured in a same trap, we make a distinction between the filling efficiency $E_{fill}$ \textit{i.e.} the proportion of occupied traps by at least one particle: 
\begin{equation}
    E_{fill} = \dfrac{number \ of \ occupied \ traps}{number  \ of \ traps} 
\end{equation}
and the capture efficiency $E_{capt}$ \textit{i.e.} the proportion of trapped particles: 
\begin{equation}
    E_{capt} = \dfrac{number \ of \ trapped \ particles}{number \ of \ particles} 
\end{equation}
Since the number of particles ($\sim 2 \rm{\ number \ of \ traps}$) is greater than the number of traps, we can expect that $ E_{fill} >  E_{capt}$. In order to optimise these objectives, several dimensionless geometric parameters, defined arbitrarily by us, have been varied:
 
 \begin{itemize}
     \item The centering $C$ represents the shift between the inlet and outlet channels in the lateral direction and thus allows studying symmetry/asymmetry effects between the inlet and the outlet. Its value is $100 \% $ when the inlet and the outlet were perfectly aligned (as in Figure \ref{fig_geom}) and $0 \%$ when the shift is maximal. $C$ is given by:
     \begin{equation}
          C = 1 - \frac{\Delta y_{io}}{\Delta y_{io \ max}} = 1-\frac{2 \ \Delta y_{io}}{W-w_{io}}
     \end{equation}
     where $\Delta y_{io}$ is the lateral distance between the inlet and outlet and $\Delta y_{io \ max}$ is its maximal value.
     
     \item The width ratio  $W_r$ in the lateral direction between the trap array and the cavity width is defined by: 
     \begin{equation}
         W_r = \frac{w}{W}
     \end{equation}
    This ratio allows to investigate the role of a particle bypass on the sides of the cavity.
  
     \item The length ratio $L_r$ in the stream direction between the trap array and the cavity length is defined by:
     \begin{equation}
         L_r = \frac{l}{L}
     \end{equation}
      This ratio allows investigating possible entrance effect of particle distribution.

     \item The trap array aspect ratio $A_r$ is defined by the ratio between the number of lines $N_l$ in stream direction and the number of columns $N_c$ in lateral direction of the trap array: 
     \begin{equation}
         A_r = \frac{N_l}{N_c}
     \end{equation}
    This ratio is maybe the less intuitive, but we will see that it may play a role. Most studies show trapping array with an $A_r<1$ but without motivating this choice. We thus intend to inspect its role. 
    
     \item  And finally, in order to characterize the importance of the channel width occupied by particles at the inlet, we introduce, the inlet and outlet channels to cavity width ratio in the lateral direction: 
    \begin{equation}
        Wch_{r} = \frac{w_{io}}{W}
    \end{equation}
 \end{itemize}
All the trapping arrays described in the literature have a regular trap distribution in the cavity. We therefore question the role of this regularity and how the results are affected when disorder is introduced into the traps positioning. To do so, we start from a regular trap position structure (Figure \ref{fig_geom}) and introduce irregularities in the trap network. This is achieved by applying a randomized translation along the horizontal and vertical directions of each trap "$i$":

\begin{equation}
    \begin{cases}
    x_i = x_{oi} + \psi_x(i) \ D_f \ \Delta x \\
    y_i = y_{oi} + \psi_y(i) \ D_f \ \Delta y \\
    \end{cases}
    \label{eq_random_pos}
\end{equation}
where $(x_i,y_i)$ are the coordinates of the $i^{th}$ trap bottom left corner (point $M$ in Figure \ref{fig_geom}), $(x_{0i},y_{0i})$ its original coordinates following the rectangular staggered pattern, $\psi_x$ and $\psi_y$ are independent random distributions of numbers between $-1$ and $1$, $D_f$ a positive constant called "disorder factor" between $0$ and $1$. We have tested two ways of introducing disorder into the lattice, uniform and centered Gaussian with a standard deviation of $0.3$ distributions for $\psi_x$ and $\psi_y$. 
\subsection{Experimental protocol}
\begin{figure}[H]
    \centering
    \includegraphics[width=0.8\textwidth]{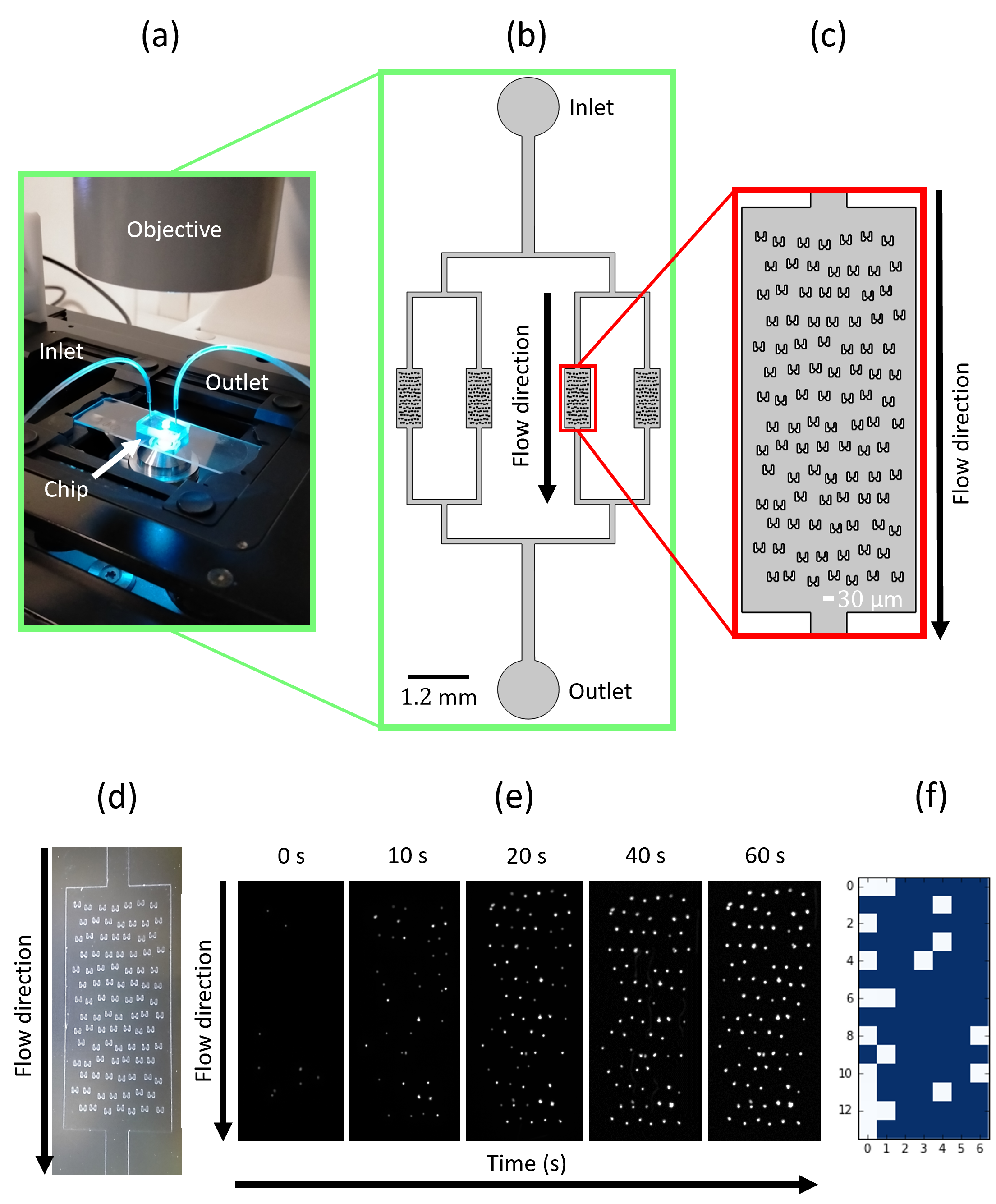}
    \caption{(a) Picture of the experimental setup. The microfluidic chip is connected to a pressure regulator with inlet and outlet tubes. The particle and fluid flow is recorded on the microscope. (b) Scheme of the microfluidic chip constituted by $4$ identical trapping arrays connected in parallel. (c) Zoom on one disordered microfluidic trapping array (obtained for $D_f=0.3$). (d) PDMS fabricated chamber following standard photolithography process observed thanks to a binocular magnifier. (e) Fluorescence microscopy images of the trapping array filling with $5$ µm polystyrene bead at different times of the experiment. (f) Binary matrix of the trapping array stationary filling. blue boxes correspond to occupied traps whilst white boxes represent empty traps.} 
    \label{fig_exp_total}
\end{figure}

The precise microfabrication steps were the same as those for \cite{Mesdjian2021} and are very briefly mentioned in this section. A microfluidic chamber resulting from disorder study is fabricated using single layer standard photo-lithography techniques. Based on previous trapping array studies in the literature, we have chosen to connect $4$ identical disordered trapping arrays in parallel \cite{Skelley2009} (Figure \ref{fig_exp_total}b,c). A chrome photomask is fabricated by direct laser writing. A circular $h=14$ µm  thick layer of photoresist resin is generated on a silicon wafer following a spin coating protocol. A  master is obtained by UV-light activation and dissolution of the non-activated photoresist resin. PDMS is mold on the master and the microfluidic chip is closed by sticking a glass coverslip on the chip following a plasma activation protocol. The microfluidic chip is then observed using fluorescent microscopy with a 10X magnification  (Figure \ref{fig_exp_total}d). The inlet and outlet channels are connected to a pressure controller with a pressure drop $ \Delta p $ corresponding to a flow rate $Q_v = 0.5$ µl min$^{-1}$ and a $ C_p = 10^6$ ml$^{-1}$ concentrated solution of fluorescent particles (Figure \ref{fig_exp_total}a), hydraulic resistance variation of the chip is negligible during all the experiment \cite{Mesdjian2021}.

\section{Results and discussion}
\label{section_Results}
\subsection{Parametric study of cavity geometry}
\begin{figure}[H]
    \centering
    \includegraphics[width=\textwidth]{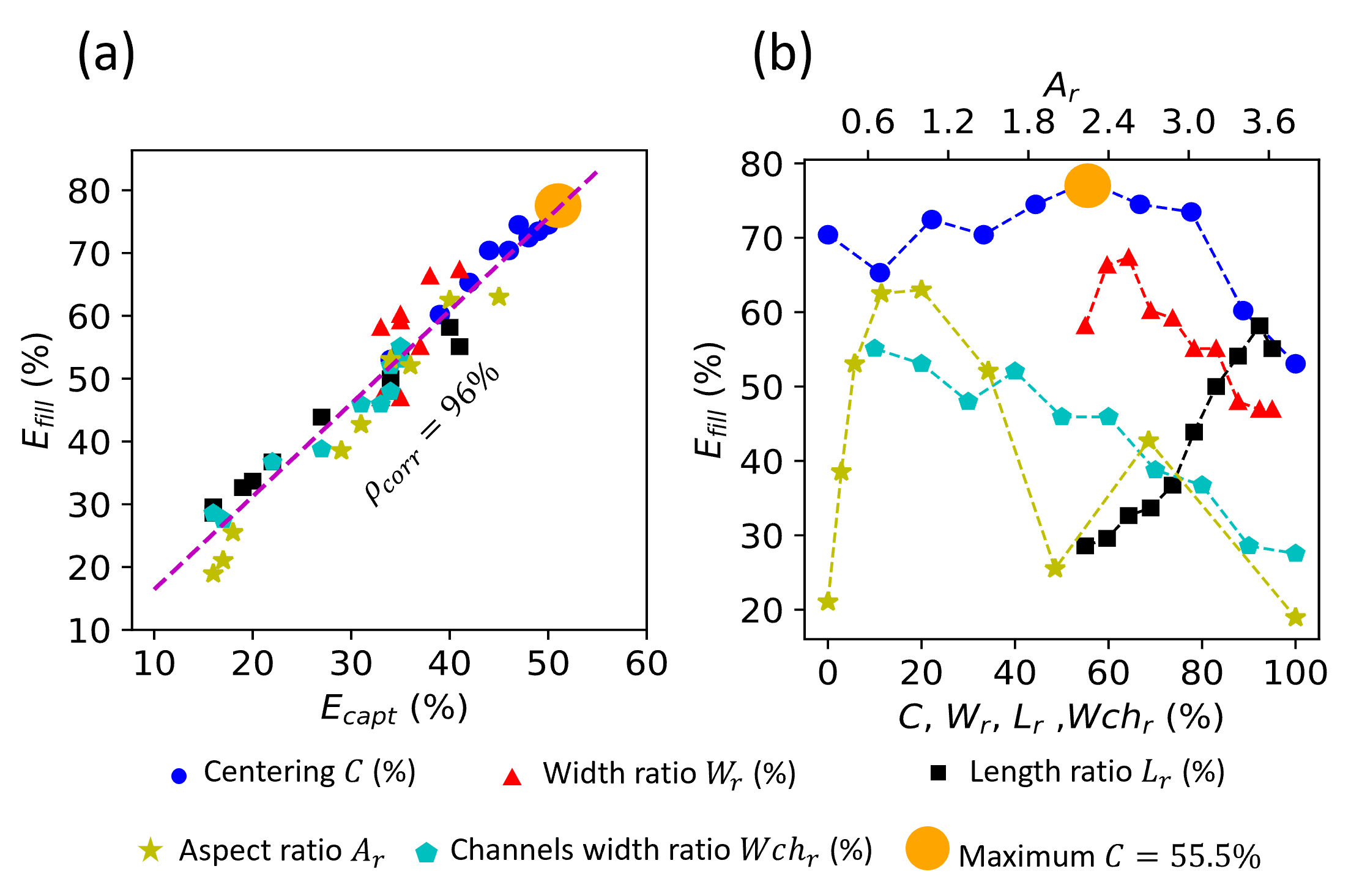}
    \caption{(a) Filling efficiency as a function of capture efficiency for each parametric study, the linear correlation coefficient is $\rho_{corr} = 96\%$. (b) Influence of the five cavity dimensionless parameters $C$, $W_r$, $L_r$, $A_r$ and $Wch_r$ on filling efficiency.}
    \label{Fig_influence_param}
\end{figure}

We study the influence of the five cavity dimensionless parameters: \(C\), \(W_r\), \(L_r\), \(A_r\), and \(Wch_r\) on filling efficiency \(E_{fill}\) and capture efficiency \(E_{capt}\) as defined in Section \ref{sec_MM_param}. Each parameter’s influence is screened over ten values while keeping the other parameters constant. In Figure \ref{Fig_influence_param}a, filling and capture efficiencies are shown to be 96\% correlated over the 50 simulations conducted. Thus, optimising the trapping array geometry using \(E_{fill}\) or \(E_{capt}\) as the objective function should yield similar trends and optimal parameter values. Therefore, the study focuses solely on filling efficiency, with capture efficiency results provided in supplementary Figure \ref{Fig_param_capt} in Section \ref{sec_SI_param}. Figure \ref{Fig_influence_param}b summarizes the evolution of filling efficiency for a regular trapping array, indicating that all cavity parameters influence \(E_{fill}\), with \(W_r\) showing an impact of ±20\% and \(A_r\) ±44\%.  The maximal filling efficiency is obtained for $C=55.5$\% (orange point).

\begin{figure}[H]
    \centering
    \includegraphics[]{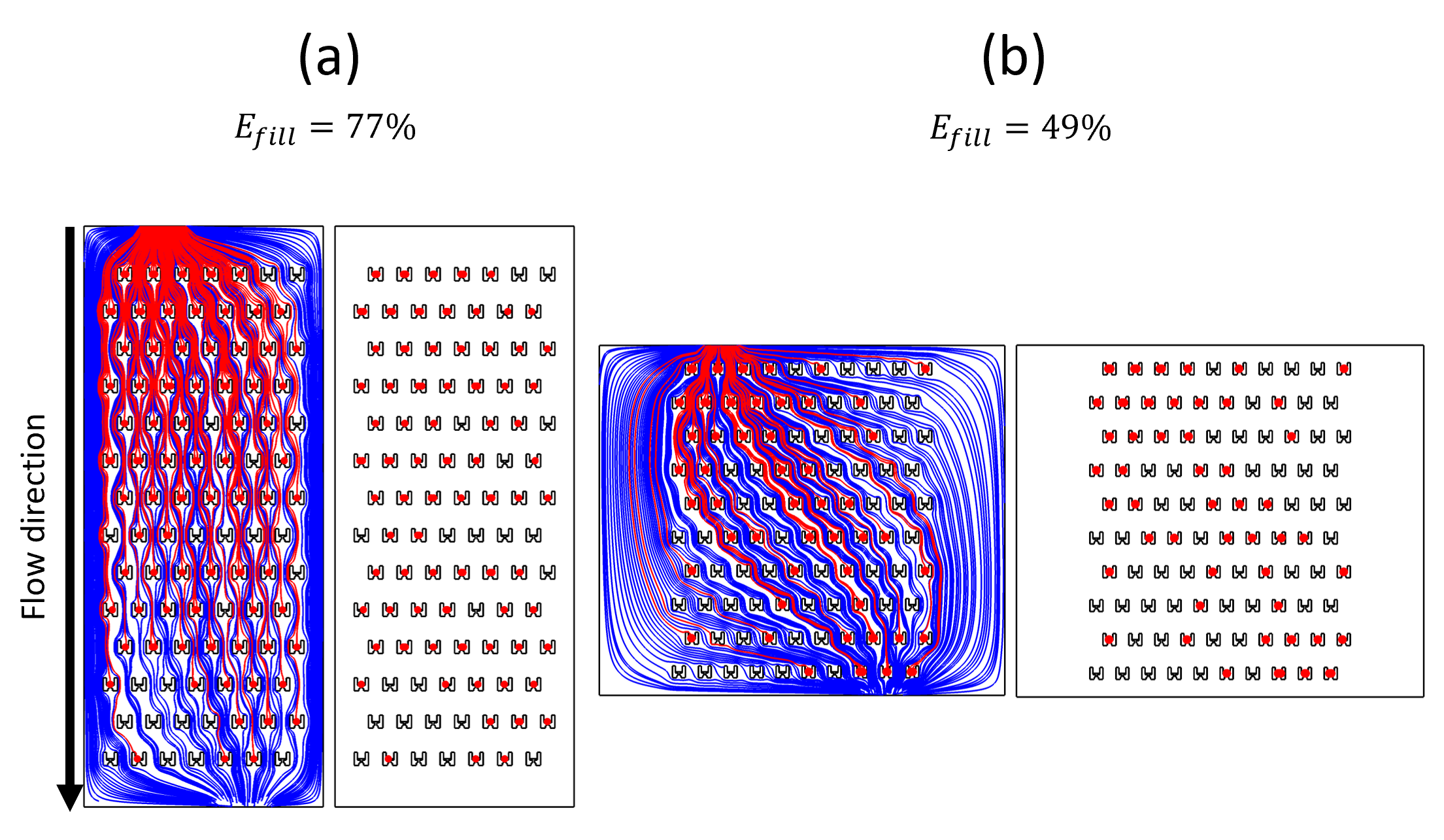}
    \caption{(a) Particle trajectories and final positions for the best geometry resulting from single parametric study. (b) Particle trajectories and final positions for the geometry obtained combining each optimum values from the each parametric study. The corresponding parameter values are: $C=55.5\%$, $W_r = 64\% $, $L_r = 90\%$, $A_r = 1$ and $Wch_r = 10\%$. The trapped particles and their trajectories are represented in red, the untrapped particle trajectories are represented in blue. The trapped particle diameter is increased to improve readability.}
    \label{fig_traj_pram_simple}
\end{figure}

Figure \ref{fig_traj_pram_simple}a shows the particle trajectories and final positions in the best geometry we get so far (corresponding to the orange point previous Figure \ref{Fig_influence_param}b). We see that trapped particle trajectories exhibit zigzag-like patterns between traps that are slightly deflected in the lateral direction (from the left to the right, red trajectories Figure \ref{fig_traj_pram_simple}a). We previously highlighted the performances of such oblique flow in the case of a maximal inlet/outlet shift ($C=0 \% $) \cite{Mesdjian2021}, but it seems that an optimum inlet/outlet shift exists for a relative centering $C$ between $44\%$ and $66\%$ (Figure \ref{Fig_influence_param}b).
Nonetheless, an optimal inlet/outlet shift is not the same for a different inlet/outlet channel width for example (different $Wch_r$). Thus, a simple optimisation approach fixing each parameter to its optimal value from this first study can not result in an efficient geometry. Indeed, the geometry built with each optimum values taken separately exhibits a low filling efficiency ($49\%$) with respect to the current optimal one ($77\% $), see Figure \ref{fig_traj_pram_simple}b. We deduce that optimisation of trapping array-chamber trapping efficiency is a complex and non-intuitive process and is about finding a compromise between combinations of geometric parameters which justifies a multiparametric optimisation approach.

\subsection{Optimisation of a regular trap array}
\label{sec_resultats_opti}
\begin{figure}[H]
    \centering
    \includegraphics[]{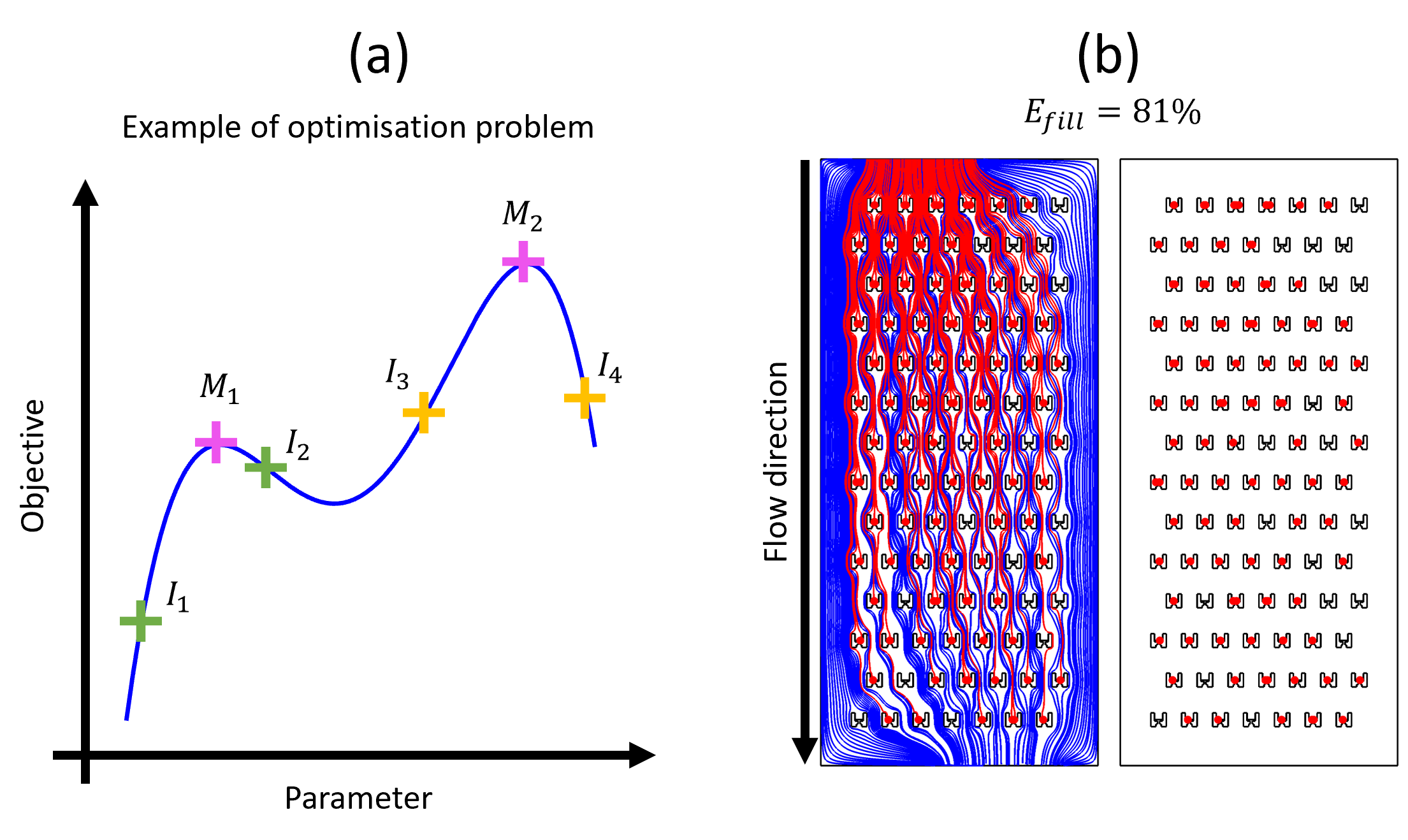}
    \caption{(a) Sketch of the local optimisation principle for a single parameter objective function. $M_1$ and $M_2$ are maximums of the function.
    $I_1$, $I_2$, $I_3$ and $I_4$ points represent $4$ different choices of initial values. (b) Particle trajectories and final positions in the globally optimised chamber. The trapped particles and their trajectories are represented in red, the untrapped particle trajectories are represented in blue. The trapped particle diameter is increased to improve readability.}
    \label{traj_opti}
\end{figure}

This section explores the use of local (Nelder-Mead \cite{NelderMead}) and global (Monte Carlo \cite{ComsolDoc}) optimisation techniques in maximizing filling efficiency \(E_{fill}\), as defined in Section \ref{sec_resultats_opti}. The optimisation problem consists in finding optimal set of parameter values leading to a maximum of an objective function. For implementation simplicity, we defined the optimisation objective function as capture efficiency: $E_{capt}(C, W_r,L_r,A_r,Wch_r)$. The optimisation results should also be optimal in filling efficiency due to the strong correlation between both metrics.
Figure \ref{traj_opti}a illustrates the local optimisation technique for a single variable objective function (blue curve) having two maxima: $M_1$ and $M_2$ points. $M_1$ is a local but not global maximum of the objective function whilst $M_2$ is the global maximum of the objective function. First, an initial set of parameter value is chosen and the objective is calculated: $I_1$, $I_2$, $I_3$ and $I_4$ represent $4$ different choices of initial values. A local optimisation algorithm follows the fastest growth of the objective function from the chosen initial point with or without calculating the derivative(s) of the objective function like a blinded walker trying to reach a peak of a mountain chain \cite{methodes_opti}. Since we have no proof of $E_{capt}$ differentiability with respect to the parameters, we choose a derivative-free method: the simplex-based Nelder-Mead method. Depending on the initial parameter values, the local optimisation algorithm converges towards different maxima: from the $I_1$ and $I_2$ points, the convergence point is $M_1$ whilst from the $I_3$ and $I_4$ points, the convergence point is $M_2$. The main advantage of local optimisation is the low number of model evaluations needed to find an optimum. However, it can stick into local maxima and miss the global maximum.

 All the different local optimisations we tried converge towards different local maxima. Thus, we used a global optimisation algorithm (Monte Carlo) which can not stick into local maxima. This technique randomly samples points within a uniform distribution inside the domain of parameters variations specified by us \cite{methodes_opti}. The convergence of such algorithm is low and not guaranteed. However, since the model has a low computational cost, this method seems appropriate and we limit the number of model evaluation to $2000$ for approximately $1$ day of computation time. As expected, we obtain our best filling efficiency with this algorithm: $E_{fill}= 81 \%$, which is $4 \%$ better than the best geometry obtained by the parametric study (with $E_{fill}=77 \% $).

Figure \ref{traj_opti}b shows the particle trajectories and final positions for  the globally optimised geometry. The corresponding parameter values are: $C= 44\% $, $W_r=78\% $, $L_r = 87\%$, $A_r = 50 \%$, $Wch_r= 48\%$. Interestingly, it seems difficult to reach a filling efficiency better than about $81\%$. In addition, the fact that $C \neq 100\%$ confirms the importance of an oblique flow that allows increasing filling efficiency. We believe that the symmetry breaking between the upstream and downstream flow allows a lateral dispersion of the particles. As such, mass transport in the lateral direction of the flow should be improved by adding disorder in the design of the trapping array which is the subject of the next section.

\subsection{Improving filling efficiency using disorder} 
\label{sec_desordre}
Although the numerical model has been validated previously \cite{Mesdjian2021}, we conducted a preliminary experimental verification to assess the efficiency of a disordered trapping array. This experiment utilizes $5$~µm diameter fluorescent polystyrene beads within a uniformly disordered array with a disorder factor of $D_f = 0.3$. Under controlled flow conditions of $0.5$~µl min$^{-1}$, a steady-state filling is attained within $60$~s, yielding a filling efficiency of $81\%$ (Figure \ref{fig_exp_total}e,f). This efficiency markedly surpasses the typical $50\%$ filling observed in symmetric chambers \cite{Mesdjian2021}. This result highlight the impact of disorder on filling efficiency, motivating further exploration through detailed numerical simulations.

\begin{figure}[H]
    \centering
    \includegraphics[width =\textwidth]{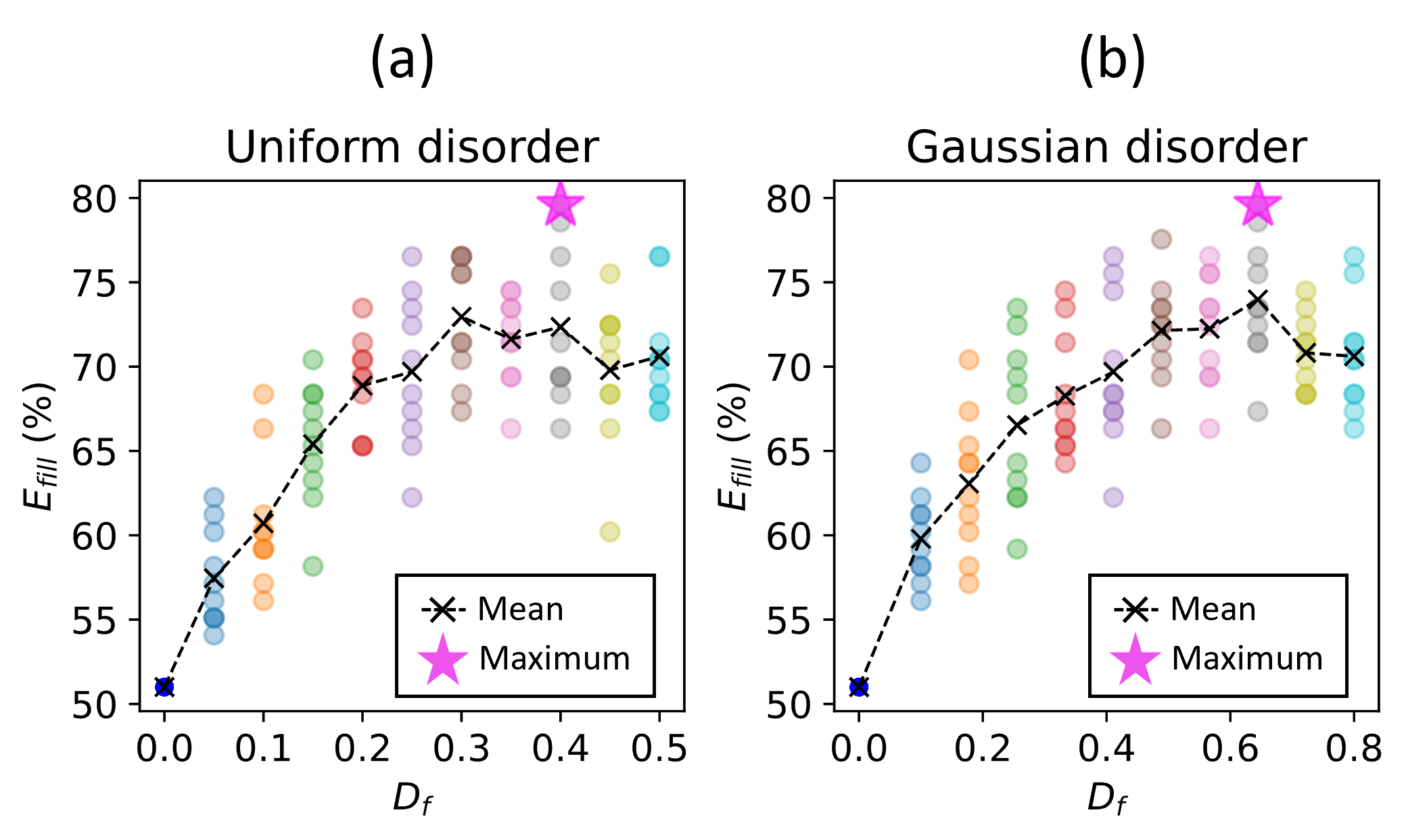}
    \caption{Evolution of filling efficiency in a disordered trapping array with a symmetric cavity (same cavity as in Figure \ref{fig_geom}) with respect to a uniform (a) and a Gaussian distribution (b) of the disorder factor $D_f$. Each simulation is represented by a dot. For each $D_f$ value, the geometry is re-built $10$ times, which corresponds to a fixed colour of vertically aligned dots. Dot colour appears darker when several point overlaps \textit{i.e} when several simulation results are equal in filling efficiency for a same $D_f$ value. Black crosses represent the mean of $E_{fill}$ values for each $D_f$ value. Black crosses are connected by dashed lines to improve visibility.}
    \label{Fig_graph_desordre}
\end{figure}

Therefore, we now investigate how the addition of disorder on the trap positioning may affect filling efficiency by numerical simulations. For this purpose, we explore geometries generated as described in section \ref{sec_MM_param}. The initial geometry is the reference geometry (shown Figure \ref{fig_geom}) of low filling efficiency  ($E_{fill} = 51\%$). Disorder is then added in the trapping array using $10$ different values of the disorder factor $D_f$ introduced in section \ref{sec_MM_param}. As the trapping array geometry is built with probability laws, the repeatability of the results is assessed by rebuilding the geometry $10$ times for each $D_f$ value. 
Figure \ref{Fig_graph_desordre} shows the evolution of $E_{fill}$ with respect to the disorder factor $D_f$ for uniform (a) and Gaussian (b) disorders. We notice that for a same $D_f$ value (same colour), a variety of different geometries can be obtained, leading to a dispersity of filling efficiency about $15\%$ (bounded by the maximum and minimum values). For both disorder distributions, we observe that the mean filling efficiency increases with the disorder parameter and reach a plateau about $\sim 72 \%$ (black crosses) at $D_f\geq 0.3$ for the uniform and at $D_f\geq 0.5$ for the Gaussian disorder.
For both disorder distributions, a maximum of filling efficiency is identified to be of $80\%$ (pink stars) which is very close to the previous one of $81 \%$ obtained for the optimisation of a regular trapping array. The best geometries are respectively obtained for $D_f= 0.40 $ and $D_f = 0.64 $ for uniform and Gaussian distributions respectively. Therefore, uniform and Gaussian disorders are equally efficient and can reach very similar efficiencies as an optimised oblique chamber with a staggered trap array.

\begin{figure}[H]
    \centering
    \includegraphics[]{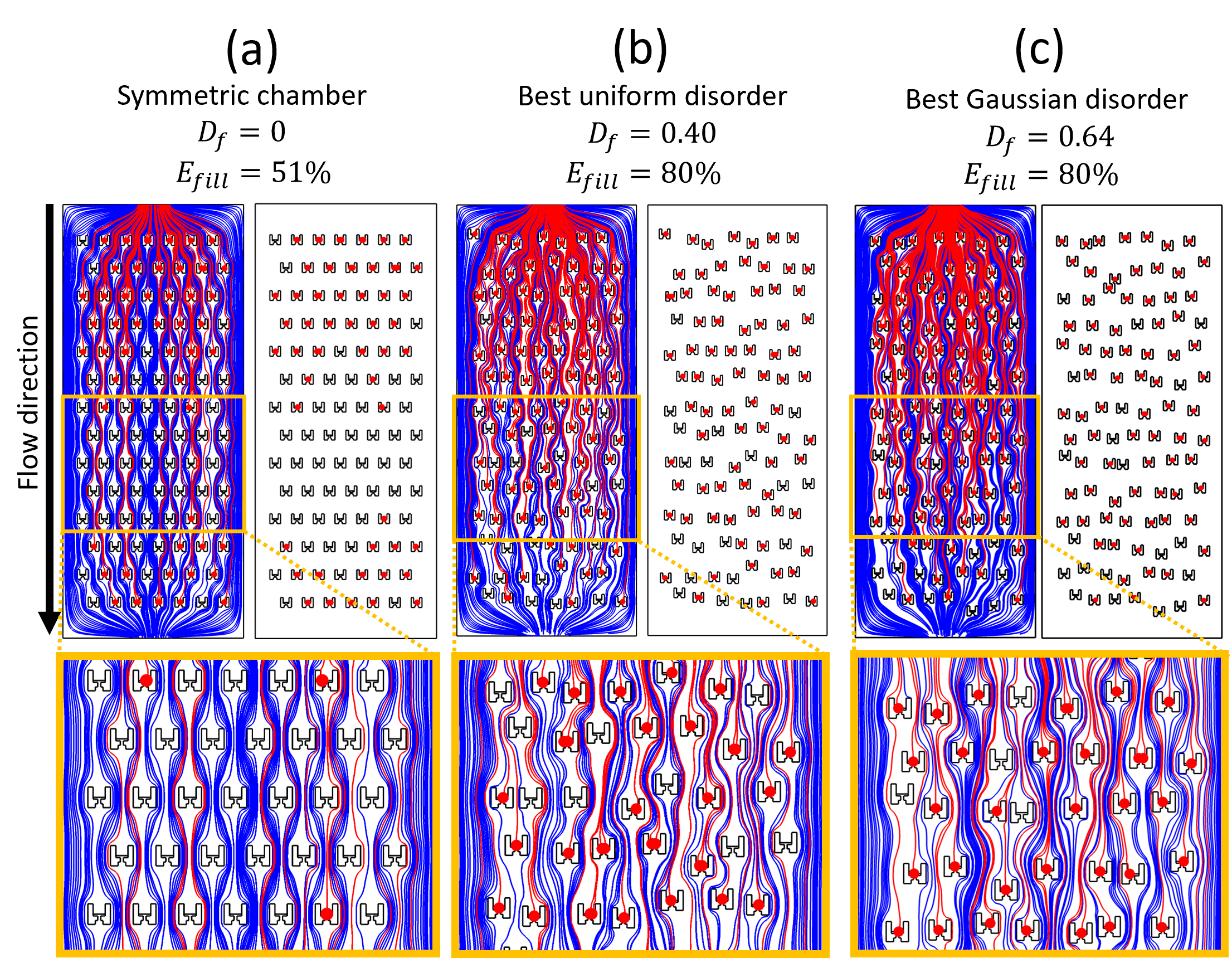}
    \caption{Particle trajectories and final positions in a symmetric chamber (a), the disordered chamber for the best uniform (b) and Gaussian (c) distributions. The trapped particles and their trajectories are represented in red, the untrapped particle trajectories are represented in blue. The trapped particle diameter is increased to improve readability.}
    \label{Fig_traj_desordre}
\end{figure}

Figure \ref{Fig_traj_desordre} shows the particle trajectories and final positions for the symmetric reference chamber (a), the best uniform (b) and best Gaussian (c) distributions of the disorder factor. The respective filling efficiencies are $51\% $ (reference geometry) and $ 80\% $ for both uniform and Gaussian disorders (corresponding to the big pink stars Figure \ref{Fig_graph_desordre}a,b). 

In a symmetric chamber where inlet and outlet channels are perfectly aligned, streamlines exhibit upstream/downstream symmetry along the traps far enough from the inlet and outlet (Figure \ref{Fig_traj_desordre}a). Therefore, the first trap rows are filled by the particles in the flow direction and almost all the next trap rows stay empty. Then, non-trapped particles display zigzag-like trajectories around the traps, see zoom of Figure \ref{Fig_traj_desordre}a. Breaking the upstream/downstream flow symmetry thanks to the addition of disorder modify the symmetric streamline patterns and allows a spanwise mass transport up to eventually filling a next trap. For optimised oblique trapping arrays, this spanwise mass transport is provided by the shift between the inlet and the outlet channels. A symmetry breaking of upstream/downstream streamlines in disordered trapping array is also clearly visible in the zooms of Figure \ref{Fig_traj_desordre}b and \ref{Fig_traj_desordre}c. Indeed, a saddle point appears on the upstream side of the trap, and streamlines recombine downstream. By breaking the symmetry, streamlines may explore traps in the lateral direction, a possibility that would not have been possible with a staggered trap array. Therefore, we think that lateral dispersion is a key mechanism allowing flow symmetry breaking and then particle trapping. Such mechanism should be quantitatively investigated which is the subject of the next section.
Before this quantitative analysis, we would like to point a remark: the significant increase of filling efficiency is not observed when adding disorder to the optimised oblique cavity presented in previous section (see supplementary section \ref{sec_desordre_sur_opti}). In this case, filling efficiency stagnates in average and the best filling efficiency is $78\% $ (see supplementary Figure \ref{fig_desordre_sur_opti}a), which is close to the best filling efficiencies obtained so far of $80 \%$, Figure \ref{Fig_traj_desordre}b). Therefore, we think that uniform disorder allows to maintain approximately $80 \%$ of filling efficiency independently of the cavity geometry. We can take advantage of this independence to increase the size of the trap array in the lateral direction at will (see supplementary section \ref{sec_reseau_geant}).

\subsection{How lateral dispersion influences trapping efficiency}
\label{sec_dispersion}

Particle pair dispersion in the lateral direction is quantified by the lateral distance between two particles $d_{\perp}(t)= y_2(t)-y_1(t)$ where $y_1(t)$ and $y_2(t)$ are the lateral coordinates of two initially close particles. At the whole population scale, lateral pair dispersion is characterized by the mean-square lateral distance $\langle d_{\perp}^2(t) \rangle$ and the mean-square deviation $\sigma_\perp^2(t) = \langle (d_{\perp}(t) -\langle d_{\perp} \rangle (t))^2 \rangle = \langle d_{\perp}^2(t) \rangle - \langle d_{\perp}(t) \rangle^2  $ of the lateral particle pair distance $d_{\perp}(t)$. 

\begin{figure}[H]
    \centering
    \includegraphics[]{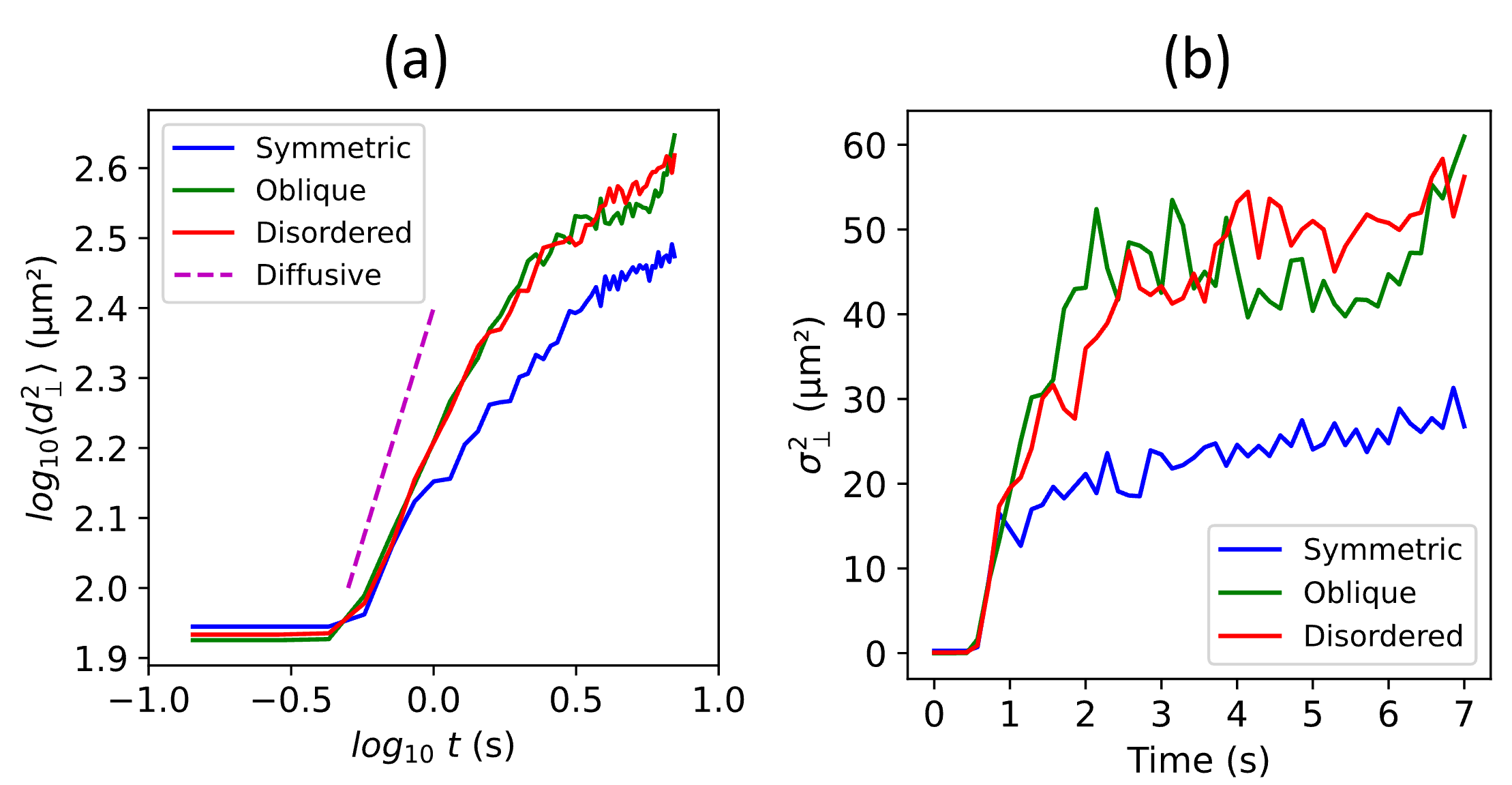}
    \caption{Statistics of lateral dispersion in the three chambers: symmetric, oblique and disordered. (a) Logarithm of the mean square lateral distance $ \langle d_{\perp}^2 \rangle$ as a function of the logarithm of time $t$. The "Diffusive" line has a unitary slope and  (b) Mean-square deviation $\sigma_\perp^2$ of the lateral distance $d_{\perp}$ between particles as a function of time $t$.}
    \label{fig_dispersion}
\end{figure}

Figure \ref{fig_dispersion}a presents the time variations of $\langle d_{\perp}^2 \rangle (t)$ on a $\text{log}_{10}$-$\text{log}_{10}$ scale. In diffusive transport, particle pair distance scales proportionally with time:  $\langle d_{\perp}^2 \rangle (t) \propto t $, exhibiting a unitary slope (pink dashed line) in Figure \ref{fig_dispersion}a. In microfluidic trapping arrays, advection dominates over diffusion, resulting in a sub-diffusive dispersion regime (see Figure \ref{fig_dispersion}a) characteristic of microfluidic flows in porous media \cite{De_anna}. The trends in Figure \ref{fig_dispersion}a indicate two distinct phases for all trapping array geometries: an initial quasi-linear growth with a slope slightly below that of diffusion and a subsequent slowdown phase. For the symmetric chamber, the linear phase is brief (blue curve) and transitions rapidly to the slowdown phase (blue curve after $\text{log}_{10}t=0$~s). In contrast, for the disordered and optimised oblique chambers, the linear regime persists longer (red and green curves until $\text{log}_{10}t \approx 0.3$~s), leading to faster lateral dispersion.

Figure \ref{fig_dispersion}b displays the time evolution of the lateral mean-square deviation $\sigma_\perp^2(t)$. interestingly, it seems to reach an asymptotic value for the symmetric chamber at the larger times (blue curve) while the values continue to increase for the disordered and oblique chambers (red and green curves). Moreover, we still note a brief quasi-linear growth of $\sigma_\perp^2(t)$ ending at $t = 1$~s (\textit{i.e} $\text{log}_{10}t=0$~s Figure \ref{fig_dispersion}a). For the disordered and optimised oblique chambers, this phase extends to $t = 2$~s, (\textit{i.e} at $\text{log}_{10}t=0.3$~s Figure \ref{fig_dispersion}a) which results in a $\sigma_\perp^2(t)$ twice larger than for the symmetric chamber. The trends and values of $\sigma_\perp^2(t)$ are similar for the disordered and optimised oblique chambers.

These observations confirm the hypothesis that lateral dispersion plays a critical role in improving filling efficiency. While the optimised oblique and disordered chambers achieve lateral dispersion through different mechanisms, their dispersion regimes are similar, which explains their comparable filling efficiencies $(\approx 80\%)$. In contrast, the symmetric chamber, with significantly lower lateral dispersion, exhibits poorer filling efficiency, ($\approx 50\%$).

\newpage
\section{Conclusion}
\label{section_conclusion}

In this study, we investigate the trapping efficiency of dense single layer trapping arrays. We adapt our previously validated CFD/particle model to parametrically optimise the filling efficiency with five dimensionless geometric parameters. After confirming that each parameter influences the capture and filling efficiencies, we use different algorithmic optimisation approaches (local and global) on a regular trapping array. The results show the importance of creating an oblique flow to increase filling efficiency. These oblique flows break the upstream/downstream flow symmetry around the traps and thus enhance lateral dispersion of the particles favouring particle exploration over a wider area. While improving lateral dispersion is a critical point in favouring trap filling, we show that similar filling efficiencies can be obtained by introducing disorder into the trap paving. An advantage of this configuration is that is does not constrain the cavity and inlet/outlet geometry for similar filling efficiencies than an optimised oblique flow geometry. Finally, a quantitative study of lateral dispersion in the optimised oblique, disordered and symmetric geometries reveals the link between the lateral dispersion of particles an trapping array filling efficiency. We think that further experimental investigations on very large trapping arrays should finally reveal which geometry is the most interesting.

\section{Acknowledgements}
This work has received support from the administrative and technological staff of “Institut Pierre-Gilles de Gennes” (Laboratoire d’excellence : ANR-10-LABX-0031, “Investissements d’avenir” : ANR-10-IDEX-0001-02 PSL and Equipement d’excellence : ANR-10-EQPX-0034). NR acknowledges funding from the École Normale Supérieure de Rennes (ENS Rennes, Contrat Doctoral Spécifique Normalien) for PhD scholarship. MCJ acknowledges CNRS and Université de Rennes for financial support.

\section{Author contributions}
\label{section_contributions}
Concepts were proposed by MCJ and JF. Data curation was done by NR and formal analysis by MCJ and NR. Simulations were performed by NR, experiments were conducted by GF (fabrication, trapping) and NR (trapping). NR, MCJ and JF built the methodology and interpreted the results. The manuscript was written by NR and reviewed for scientific and technical aspects by MCJ and JF and for formal aspects by MCJ and RA. Work were supervised by MCJ, JF and RA. Access to simulation material was supported by RA. 

\section{Conflicts of interest}
There are no conflicts of interest to declare. 

\bibliographystyle{unsrtnat}
\bibliography{article.bib}

\newpage
\section{Supplementary information to: Materials and Method}
\subsection{Model parameters values}
\label{sec_SI_param_values}
The numerical parameters used in the model are reported in the following Table \ref{tableau_all_param}.
\begin{table}[H]
\begin{footnotesize}
     \begin{center}
     {\setlength\arrayrulewidth{1pt}     \begin{tabular}{c c c}
      \hline
      Fixed geometric parameters & Description & Value\\
      \hline\\
      $h$ & chamber's out-of-plane depth  & $14$ µm \\
      $w_o$ & Traps small opening width & $4$ µm \\  
      $w_{pil}$ & Trap pillar width & $7$ µm \\
      $w_{sr}$ & Traps small rectangle width & $10$ µm\\  
      $w_{trap}$ & Traps width & $30$ µm\\ 
      $\Delta x$ & Horizontal distance between traps  & $50$ µm\\  
      $\Delta y$ & Vertical distance between traps & $30$ µm\\ 
      $l_{sr}$ & Trap small rectangle's length & $5$ µm\\  
      $l_{trap}$ & Trap's length & $27$ µm\\ 
      $N_{traps}$ & Number of traps in the MTA & $ 96-98 $\\\\
      \hline
      Physical parameters & Description & Values \\
      \hline\\
      $\mu $ & Fluid dynamic viscosity & $0.001$ Pa s\\  
      $N_{part}$ & Number of particles & $2 \ N_{traps}$\\  
      $p_{in}$ & Uniform inlet pressure  & $100$ Pa\\  
      $p_{out}$ & Uniform outlet pressure  & $0$ Pa\\  
      $\rho$ & Fluid density & $1000$ kg m$^{-3}$ \\ 
      $R_p$  & Particles radius & $2.5$ µm \\
      $T_e$ & Study time & $3$ s - $32$ s\\\\ 
      \hline
      Parametric \& optimisation studies & Description & Values \\
      \hline\\
      $C$ &  Inlet/outlet centering & $ [0,100] \ \%$ \\
      $W_r$ &  Width ratio   & $ [55,95] \  \%$ \\
      $L_r$ & Length ratio  & $ [55,95] \  \%$ \\
      $A_r$ & Aspect ratio  & $0.3$,$0.4$,$0.5$,$0.7$,$1$,$1.5$,$2$,$2.7$,$3.8$ \\
      $Wch_{r}$ & Channel to chamber width ratio & $ [10,100] \ \%$ \\\\
      \hline \\
      \end{tabular}\\
      }
      \caption{The model parameters. For single parametric studies, $10$ parameters values are chosen to uniformly vary in the specified interval excepted for the $A_r$ parameter where $9$ values are chosen to have the same total number of traps $\pm 3 $ traps.}
      \label{tableau_all_param}
      \end{center}
      \end{footnotesize}
\end{table}

\subsection{Finite-Element implementation}
\label{sec_SI_FEM}
In COMSOL, the Brinkman equation is written in the classical Continuum Mechanics formulation:
\begin{equation}
\begin{cases}
    \boldsymbol{\nabla} \cdot \stress + \underline{f} = \underline{0} \\
    \boldsymbol{\nabla} \cdot \underline{v} = 0 \\
    \stress =  - p  \Id + 2  \mu  \D \\
    \D = \dfrac{1}{2}  \left(\boldsymbol{\nabla} \underline{v} + (\boldsymbol{\nabla} \underline{v})^T \right)\\
\end{cases}
\end{equation}
where $\stress$ is the Newtonian fluid Cauchy stress tensor, $\Id$ the identity tensor and $\D$  the strain rate tensor. The mixed weak form equation implemented in COMSOL is obtained by applying the scalar product of Brinkman equation with test velocity $\hat{\underline{v}}$ and by multiplying the mass-conservation (continuity) equation by test pressure $\hat{p}$ and then integration on the fluid domain $\Omega$ (grey domain Figure \ref{fig_geom}):
\begin{equation}
\begin{cases}
    \Int{\Omega}{(\boldsymbol{\nabla} \cdot \stress)  \cdot \hat{\underline{v}}} \ d\Omega + \Int{\Omega}{\underline{f} \cdot \hat{\underline{v}}} \ d\Omega = 0 \\ 
    \Int{\Omega}{\hat{p} \boldsymbol{\nabla} \cdot \underline{v}} \  d\Omega= 0\\ 
\end{cases}
\end{equation}
Then by integration by part of the first integral and application of the divergence theorem, the following weak form is implemented into COMSOL:
\begin{equation}
\begin{cases}
    \Int{\Omega}{-\stress : \boldsymbol{\nabla}\hat{\underline{v}}} \ d\Omega 
    + \OInt{\partial \Omega}{(\stress \cdot \underline{n}) \cdot  \hat{\underline{v}}} \ d\Gamma + 
    \Int{\Omega}{\underline{f} \cdot \hat{\underline{v}}} \ d\Omega = 0 \\ 
    \Int{\Omega}{\hat{p} \boldsymbol{\nabla} \cdot \underline{v}} \  d\Omega= 0\\ 
\end{cases}
\end{equation}
Where the first integral represents the virtual power of internal forces including viscous and pressure forces. This equation can be implemented into any open-source FEM code such as Freefem++ or Fenics. The second term represents the virtual power of boundary external forces not null only on the inlet and outlet boundaries where the pressure is prescribed and where $\hat{\underline{v}}$ is not null. The last term represents the virtual power of external body forces. The latter equations are discretised in space by applying the Finite-Element Method (FEM) with linear Lagrange shape functions for both pressure and velocity fields (P$1$/P$1$ elements). Even without advection term, such elements must be stabilised to satisfy Ladyzhenskaya–Babuška–Brezzi (LBB) condition. To do so, the COMSOL consistent streamline stabilisation term is added. The weak form  equation discretisation is obtained first by applying the nodal (Galerkin) interpolation for $\underline{v}$ and $p$ by their respective interpolation functions and then by replacing the test functions by all interpolation functions. The resulting linear system of equations is numerically inverted using the COMSOL MUMPS solver.

\section{Supplementary information to: Results}
\subsection{Cavity comprehensive parametric study: capture efficiency results}
\label{sec_SI_param}
\begin{figure}[H]
    \centering
    \includegraphics[width=\textwidth]{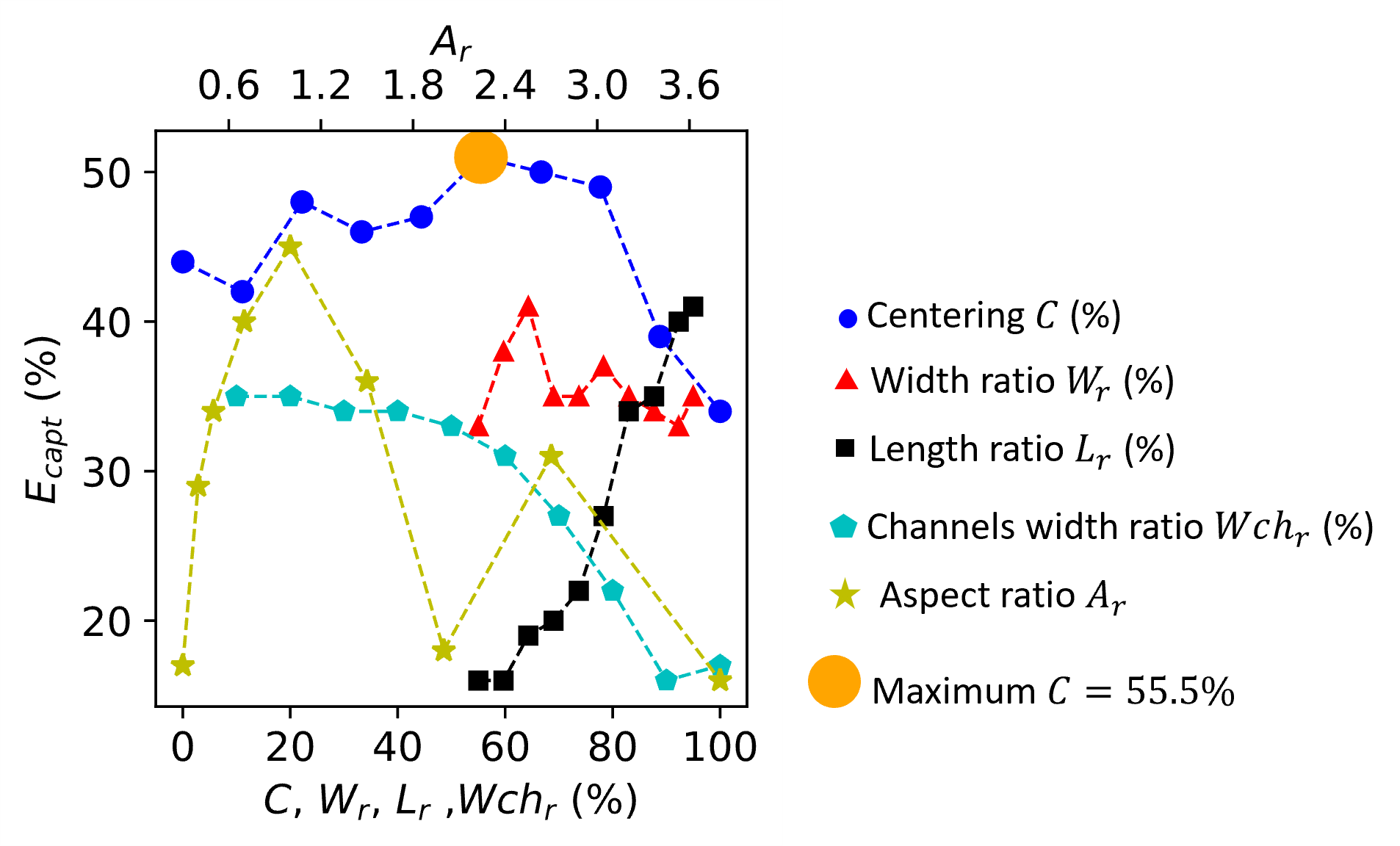}
    \caption{Influence of the five cavity dimensionless parameters $C$, $W_r$, $L_r$, $A_r$ and $Wch_r$ on capture efficiency. When parameters are fixed, their respective values are : $C=100\%$, $W_r = 84\% $, $L_r = 86\%$, $A_r = 50\%$ and $Wch_r = 23\%$ which corresponds to the reference geometry Figure \ref{fig_geom}. Results from a same parametric study are connected by dashed lines to improve readability.}
    \label{Fig_param_capt}
\end{figure}

The results of the cavity comprehensive parametric study in capture efficiency (\textit{ie} the proportion of trapped particles defined \ref{sec_MM_param}) are summarised Figure \ref{Fig_param_capt}. We notice the very similar curve trends with respect to filling efficiency displayed Figure \ref{Fig_influence_param}, leading to a $96 \%$ linear correlation coefficient between the two metrics (capture and filling efficiencies). As a consequence, the optimum geometry for both metrics is obtained for the same parameter value $C = 55.5\%$.

\subsection{Disorder parametric study for optimised cavity}
\label{sec_desordre_sur_opti}
\begin{figure}[H]
    \centering
    \includegraphics[]{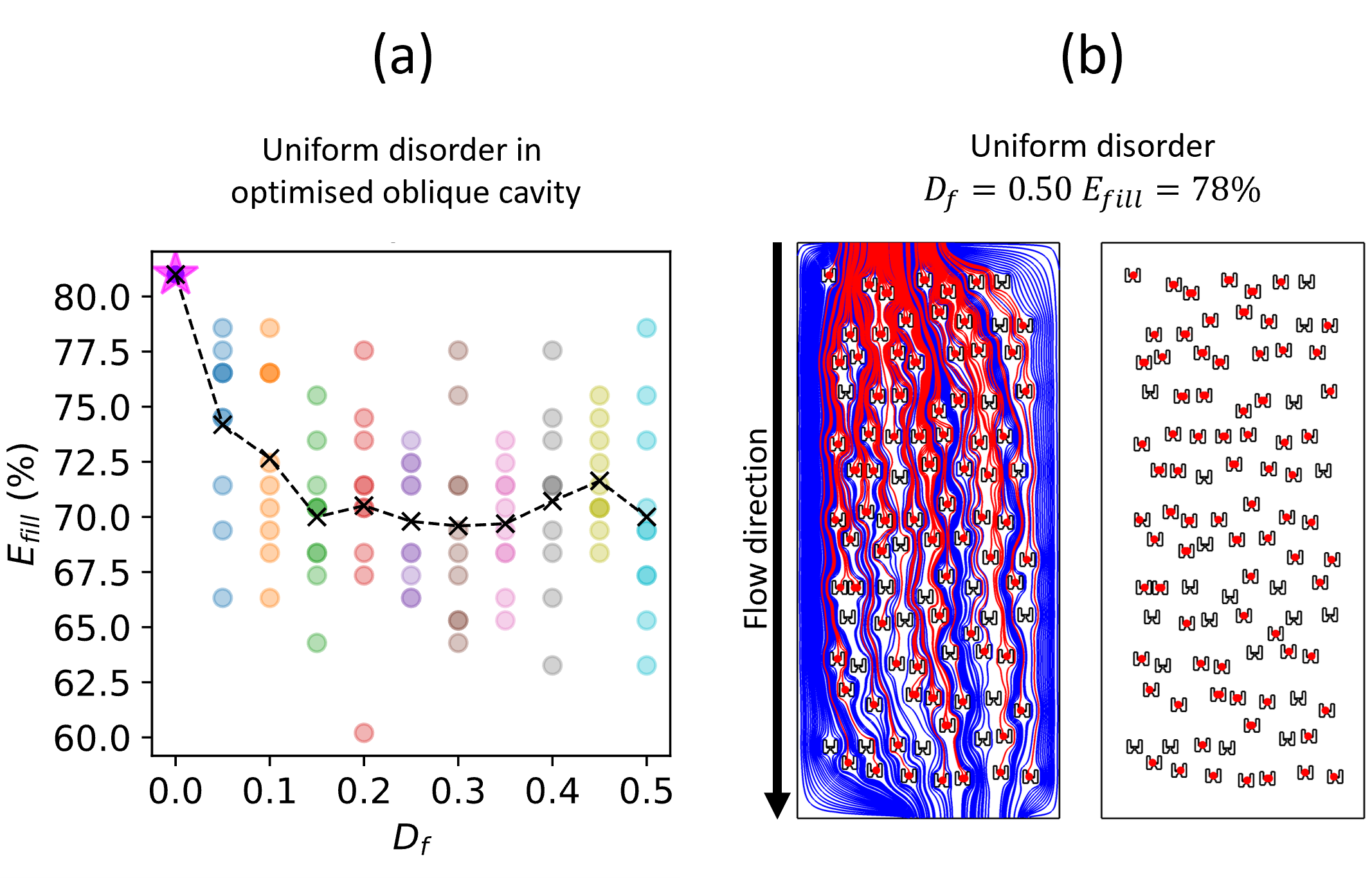}
    \caption{(a) Evolution of capture efficiency with respect to  the the disorder factor $D_f$ for a uniform disorder. Each simulation is represented by a dot. For each $D_f$ value, the geometry is re-built $10$ times, which corresponds to a fixed colour of vertically aligned dots. Dot colour appears darker when several point overlaps \textit{i.e} when several simulation results are equal in filling efficiency for a same $D_f$ value. Black crosses represent the mean of $E_{fill}$ values for each $D_f$ value. Dark crosses are connected by dashed lines to improve readability. (b) Particle trajectories and final positions for the most efficient disordered trap array with the optimised oblique cavity (corresponding to the highest point of Panel (a) with $D_f = 0.50$). The trapped particles and their trajectories are represented in red, the untrapped particle trajectories are represented in blue. The trapped particle diameter is increased to improve readability.}
    \label{fig_desordre_sur_opti}
\end{figure}

We study the influence of uniform disorder with the optimised oblique cavity presented section \ref{sec_resultats_opti} and Figure \ref{traj_opti}b. To do so, we conduct the same disorder factor $D_f$ parametric study as in section \ref{sec_desordre}: we screen $10$ values of $D_f$ with a geometry rebuilt $10$ times for each $D_f$ value. Figure \ref{fig_desordre_sur_opti}(a) shows the evolution of filling efficiency with respect to the disorder factor $D_f$. Interestingly, adding disorder slightly decreases filling efficiency because the maximum filling efficiency is obtained without adding disorder: $E_{fill} = 81\%$ and the best filling efficiency obtained with a disordered MTA is $E_{fill} = 78\%$ (highest dot for $D_f=0.5$ Figure \ref{fig_desordre_sur_opti}(a)). Therefore, the use of disordered trap array seems relevant only with a non-optimised cavity.

\subsubsection{Disordered trap arrays can be easily expanded in the lateral direction}
\label{sec_reseau_geant}
\begin{figure}[H]
    \centering
    \includegraphics{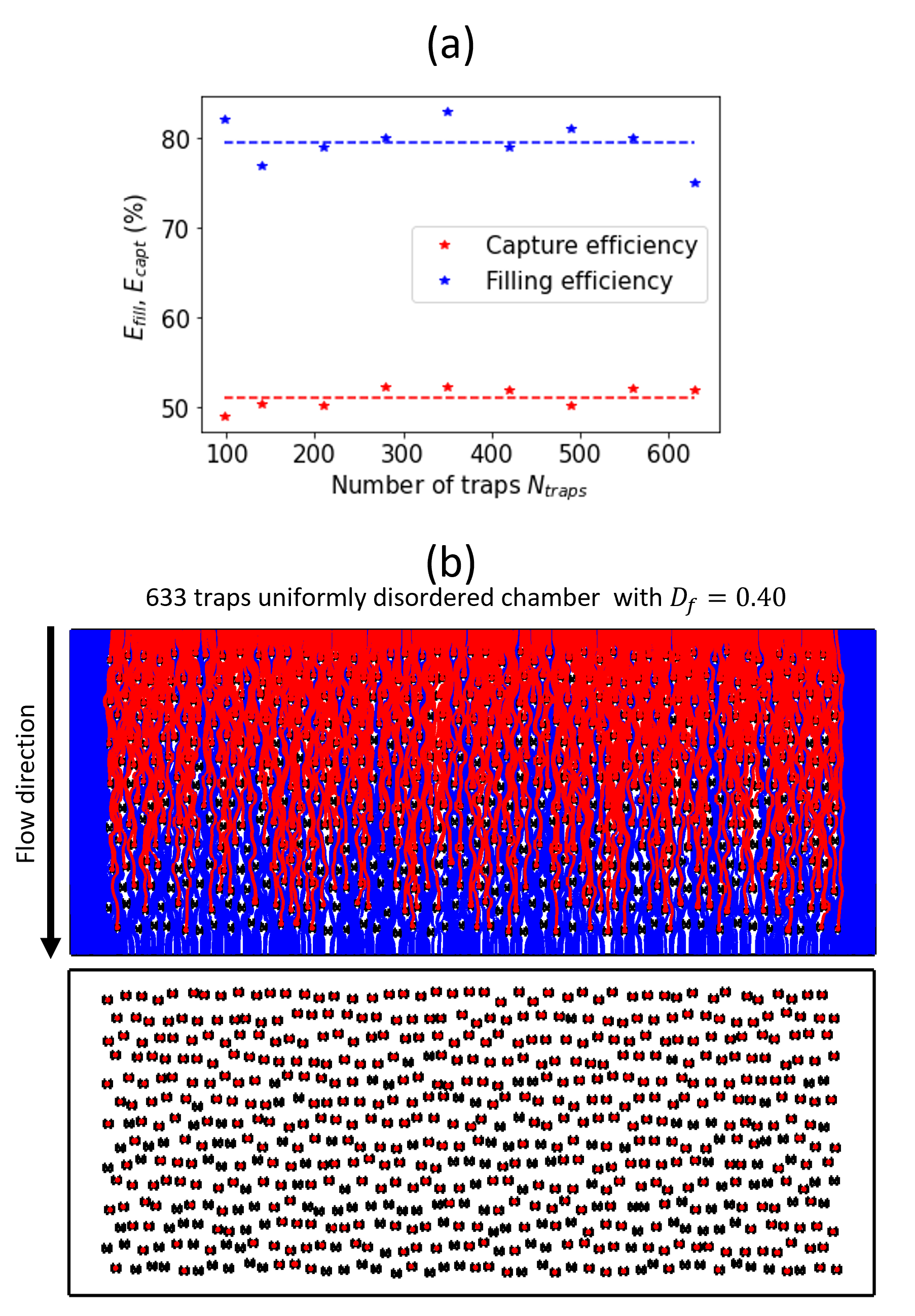}
    \caption{(a) Evolution of capture and filling efficiencies with respect to the number of traps when laterally expanding a $D_f = 0.40$ uniformly disordered MTA. Horizontal lines correspond to constant function least square fittings. (b) Particles trajectories in the biggest disordered MTA simulated on Panel (a). The trapped particles and their trajectories are represented in red, the untrapped particle trajectories are represented in blue. The trapped particle diameter is increased to improve visibility.}
    \label{fig_taille_reseau}
\end{figure}

Disordered MTAs generate lateral dispersion of particles due to the trap relative positions. This characteristic offers more possibilities of the inlet/outlet channels and cavity designs than an optimised oblique chamber where all the five dimensionless parameters are fixed. As an example, we show here that when using the largest possible inlet/outlet channels ($Wch_r = 100 \% $) and expanding the MTA size in the lateral direction, the filling and capture efficiencies are few modified and the averaged filling efficiency remains very close to $80 \%$ (Figure \ref{fig_taille_reseau}(a)) even though the MTA aspect ratio $A_r$ and width ratio $W_r$ are modified. The particle trajectories corresponding to largest chamber are shown Figure \ref{fig_taille_reseau}(b). 

\end{document}